%% file: pec_expan-bbl.tex
\documentclass{aa}

\usepackage{etoolbox}

\newtoggle{postrefAA}

\togglefalse{postrefAA} 

\newcommand\dosingle[1]{#1}  \newcommand\dodouble[1]{ } 

\newcommand\nice[1]{#1}    \newcommand\subm[1]{}   
\subm{ \renewcommand\dosingle[1]{ }   \renewcommand\dodouble[1]{#1} }

\dodouble{\documentclass[usenatbib,referee]{mnras}}

\usepackage[T1]{fontenc}
\usepackage{aecompl}

\usepackage{graphicx}

\usepackage{amsfonts}
\usepackage{amsmath}
\usepackage{rotating} 

\usepackage[varg]{txfonts}
\usepackage{fixltx2e} 

\newcommand\mystamp[1]{{}}

\newcommand\mystamppreamble{
  \usepackage{eso-pic}
  \usepackage{color}
  \definecolor{redstamp}{rgb}{0.99,0.80,0.90} 
  \usepackage{datetime}
  \usepackage[normalem]{ulem}
}

\newcommand\mystampdothestamp{
  \AddToShipoutPicture{
    \AtTextLowerLeft{
      \makebox(550,300)[c]{\resizebox{\textwidth}{!}{
          \rotatebox{36}{\textsf{\textbf{\color{redstamp}brouillon~\frdutoday~\currenttime}}}}}
    }
  }
}

\mystamp{\mystamppreamble}

\usepackage{natbib}  
\usepackage{url}

\newtoggle{prerefAA} \newtoggle{prerefBB} \newtoggle{prerefCC}

\togglefalse{prerefAA} 

\togglefalse{prerefBB} 

\togglefalse{prerefCC} 

\iftoggle{prerefAA}{
  \newcommand\prerefereeAAchanges[1]{{\bf \large \color{myred} #1}}     \usepackage{color}  \definecolor{myred}{rgb}{0.7,0.0,0.2}
}{
  \newcommand\prerefereeAAchanges[1]{#1}    
}

\iftoggle{prerefBB}{
  \newcommand\prerefereeBBchanges[1]{{\bf \large \color{myred} #1}}     \usepackage{color}  \definecolor{myred}{rgb}{0.7,0.0,0.2}
}{
  \newcommand\prerefereeBBchanges[1]{#1}    
}

\iftoggle{prerefCC}{
  \newcommand\prerefereeCCchanges[1]{{\bf \large \color{myred} #1}}     \usepackage{color}  \definecolor{myred}{rgb}{0.7,0.0,0.2}
}{
  \newcommand\prerefereeCCchanges[1]{#1}    
}

\iftoggle{postrefAA}{
  \newcommand\postrefereeAAchanges[1]{{\bf \large \color{myred} #1}}     \usepackage{color}  \definecolor{myred}{rgb}{0.7,0.0,0.2} \newcommand\postrefereeAAchangesnosize[1]{{\bf \color{myred} #1}}
}{
  \newcommand\postrefereeAAchanges[1]{#1}     \newcommand\postrefereeAAchangesnosize[1]{#1}
}

 \usepackage{hyperref}

\nice{

\providecommand{\url}[1]{\href{#1}{#1}}

}

\renewcommand{\eprint}[1]{\href{http://arxiv.org/abs/#1}{{\tt [arXiv:#1]}}}

\providecommand{\adsurl}[1]{}
\newcommand\SSS{Sect.~}

\providecommand\apj{ApJ}                 
\providecommand\apjs{ApJSupp}                 
\providecommand\apjl{ApJL}                 
\providecommand\aap{A\&A}            
\providecommand\mnras{MNRAS}

\providecommand\prd{PRD}

\providecommand\jcap{JCAP}

\providecommand\grg{Gen.~Rel.~Grav.}
\providecommand\annalesBruxelles{Ann. de la Soc. Sc. de Brux.}
\providecommand\annrevnucpartphys{Ann.~Rev.~Nucl.~Part.~Sci.}
\providecommand\SPIEConfSeries{SPIE Conf. Ser. }
\providecommand\scandjstat{Scand.~J.~Stat.}
\providecommand\annmathstat{Ann.~Math.~Stat.}
\providecommand\pnas{Proc.~Nat.~Acad.~Sci.}
\providecommand\cqg{Class.~Quant.~Gra.}   %

\providecommand\nucphysbprocsupp{Nucl.~Phys.~B~Proc.~Supp.}

\providecommand\ijmpd{Int. J. Mod. Phys. D}

\nice{  }

\providecommand\kmsMpc{\mbox{\,$\mathrm{km/s/Mpc}$}}

\newcommand\gtapprox{\,\lower.6ex\hbox{$\buildrel >\over \sim$} \, }
\newcommand\ltapprox{\,\lower.6ex\hbox{$\buildrel <\over \sim$} \, }
\newcommand\propapprox{\,\lower.6ex\hbox{$\buildrel \propto\over \sim$} \, }

\newcommand\arcs{\ifmmode {'' }\else $'' $\fi}     
\newcommand\arcm{\ifmmode {' }\else $' $\fi}       
\newcommand\ddeg{\ifmmode^\circ\else$^\circ$\fi}    

\newcommand\diffd{\mathrm{d}}

\newcommand\frtoday{Le\space\number\day\space\ifcase\month\or
  janvier\or f\'evrier\or mars\or avril\or mai\or juin\or
  juillet\or ao\^ut\or septembre\or octobre\or novembre\or
d\'ecembre\fi\space \number\year}
\def\frdutoday{du\space\number\day\space\ifcase\month\or
  janvier\or f\'evrier\or mars\or avril\or mai\or juin\or
  juillet\or ao\^ut\or septembre\or octobre\or novembre\or
d\'ecembre\fi\space \number\year}

\newcommand\todayISO{\number\year-\ifnum\month<10 0\fi\number\month-\ifnum\day<10 0\fi\number\day}

\newcommand{\CX}{{\cal X}}

\newcommand\hMpc{\mbox{$h^{-1}$\,Mpc}}

\newcommand\Omm{\Omega_{\mathrm{m}}}
\newcommand\Ommzero{\Omega_{\mathrm{m0}}}

\newcommand\Ommzeroeff{\Omega_{\mathrm{m}0}^{\mathrm{eff}}}
\newcommand\Ommzerobg{\Omega_{\mathrm{m}0}^{\mathrm{bg}}}

\newcommand\Omkzero{\Omega_{\mathrm{k}0}}

\newcommand\OmReff{\Omega_{\cal{R}}^{\mathrm{eff}}}
\newcommand\OmRzeroeff{\Omega_{{\cal R}0}^{\mathrm{eff}}}
\newcommand\OmQeff{{\Omega_{{\cal Q}}^{\mathrm{eff}}}}
\newcommand\OmQzeroeff{{\Omega_{{\cal Q}0}^{\mathrm{eff}}}}

\newcommand\aeff{a_{\mathrm{eff}}}
\newcommand\aeffzero{a_{\mathrm{eff0}}}
\newcommand\abg{a_{\mathrm{bg}}}
\newcommand\abgzero{a_{\mathrm{bg0}}}

\newcommand\dotabg{\dot{a}_{\mathrm{bg}}}
\newcommand\Hzeroeff{H_0^{\mathrm{eff}}}

\newcommand\Heff{H^{\mathrm{eff}}}

\newcommand\Hzerobg{H_0^{\mathrm{bg}}}
\newcommand\Hzerobgifbgtrue{\mbox{{$H_{1}^{\mathrm{bg}}$}}}
\newcommand\Hbg{H^{\mathrm{bg}}}

\newcommand\Hpeculiar{{H_{\mathrm{pec}}^{\mathrm{void}}}}
\newcommand\Hpeculiarzero{{H_{\mathrm{pec},0}^{\mathrm{void}}}}
\newcommand\tzeroLCDM{t_0^{\Lambda\mathrm{CDM}}}

\newcommand\tzerobg{{t_{\abg=1}}}
\newcommand\tzeroefffull{{t_{\aeff=1}}}
\newcommand\tzeroeff{{t_0}}
\newcommand\tzeroGal{t_0} 

\newcommand\RCeff{R_{\mathrm{C}}^{\mathrm{eff}}}
\newcommand\RCzeroeff{R_{\mathrm{C}0}^{\mathrm{eff}}}

\usepackage{enumitem}

\begin{document}

\title{The background Friedmannian Hubble constant in relativistic inhomogeneous cosmology and the age of the Universe}

\titlerunning{Early FLRW Hubble expansion in GR cosmology}
\authorrunning{Roukema, Mourier, Buchert \& Ostrowski}

\author{Boudewijn F. Roukema \inst{1,2} \and
  Pierre Mourier \inst{2,3},
  Thomas Buchert \inst{2},
  Jan J. Ostrowski \inst{1,2}
  }
\institute{Toru\'n Centre for Astronomy,
  Faculty of Physics, Astronomy and Informatics,
  Grudziadzka 5,
  Nicolaus Copernicus University,
  ul. Gagarina 11, 87-100 Toru\'n, Poland
  \and
  Univ Lyon, Ens de Lyon, Univ Lyon1, CNRS, Centre de Recherche Astrophysique de
  Lyon UMR5574, F--69007, Lyon, France
  \and
    {D\'epartement de Physique,
      \'Ecole normale sup\'erieure,
      24, rue Lhomond, F--75230, Paris cedex 05, France}
}


\date{\frtoday}



\newcommand\Nchainsmain{16}
\newcommand\Npergroup{four}


\abstract
  {In relativistic inhomogeneous cosmology, structure formation
  couples to average cosmological expansion. A conservative approach to
  modelling this {assumes an Einstein--de~Sitter model
  (EdS) at early times and extrapolates this forward in cosmological time
  as a ``background model''
  against which average properties of today's Universe can be measured.}}
  {This requires adopting
  {an early-epoch--normalised} background Hubble constant
    $\Hzerobgifbgtrue$.}
  {Here, we show that the $\Lambda$CDM model can be used as
  {an observational proxy to}
  estimate
  $\Hzerobgifbgtrue$ rather than choose it arbitrarily.
  We assume (i) an EdS model at early times; (ii) a zero dark
  energy parameter; (iii) bi-domain scalar averaging---division of the
  spatial sections into over- and underdense regions; and (iv)
  virialisation (stable clustering) of collapsed regions.}
  {We find
  $\Hzerobgifbgtrue= {37.7\pm0.4}${\kmsMpc} (random error only) based on
  a Planck $\Lambda$CDM observational proxy.}
  {Moreover, since the scalar-averaged expansion rate is expected to exceed
  the (extrapolated) background expansion rate, the expected age of the Universe
  should be much less than $2/(3\Hzerobgifbgtrue) = 17.3$~Gyr.
  The maximum stellar age of Galactic Bulge microlensed
  low-mass stars
  {(most likely: 14.7~Gyr;
    68\% confidence: 14.0--15.0~Gyr)} suggests an age
  about a Gyr older than the (no-backreaction) $\Lambda$CDM estimate.}


\mystamp{\mystampdothestamp}

\maketitle

\dodouble{ \clearpage } 


\newcommand\FIGWIDTHONE{0.95\columnwidth}
\newcommand\FIGWIDTHTWO{1.05\columnwidth}

\newcommand\ftmaxage{
  \begin{figure}
    \input{t_max_age}
    \caption{{Skew-normal reconstructed
        [Eq.~(\ref{e-skew-normal})]
        probability density functions $p_i$ of the twelve
      \protect\citet{Bensby13Bulgetzero} microlensed Galactic Bulge stars
      whose most likely age is greater than 13.0~Gyr
      (thin curves); and probability density function
      $\diffd P_T/\diffd t'$ of the most likely oldest age $T$ of these
      stars (thick curve),
      assuming that the stars' true ages are chosen randomly from their
      respective pdfs $p_i$
      [Eq.~(\protect\ref{e-P-T-defn})]. The vertical line indicates
      $\tzeroLCDM$.}
      \label{f-t-max-age}}
\end{figure}
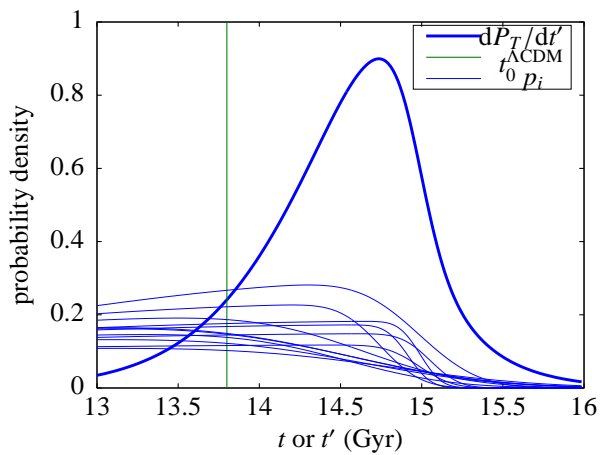} 

\section{Introduction} \label{s-intro}

The $\Lambda$CDM model, whose metric is a member of the
Friedmann--Lema\^{\i}tre--Robertson--Walker (FLRW) family, is the
standard cosmological model, but it assumes a non-standard model
of gravity.
In {other words, gravity is assumed to apply} separately to
structure formation and FLRW uniform spatial expansion, i.e.
the former is hypothesised to be gravitationally decoupled from the latter,
despite the coupling present in the Raychaudhuri equation and the Hamiltonian constraint
\prerefereeCCchanges{\citep{Buch00Hiroshima,Buch00scalav,Buch01scalav}}.
Work towards a
cosmological model in which standard general relativity determines the
relation between structure formation and expansion is incomplete
\citep[e.g.][and references therein]{EllisStoeger87,Buchert11Towards}.
A common element to many implementations
of this relativistic,
{gravitationally coupled}
approach to cosmology
is to assume an
Einstein--de~Sitter model (EdS) at early times, when density perturbations
are weak, and extrapolate this forward in cosmological time as a
``background'' model,
{adopting the same time foliation for an effective model}
{that includes gravitational coupling.}
Here, we argue that the Hubble constant
{$\Hzerobgifbgtrue$ needed to normalise
  this background EdS model at early epochs, such that the
  present effective scale factor is unity,}
cannot be chosen arbitrarily,
  since it is observationally constrained.
  The value of $\Hzerobgifbgtrue$ will be needed,
  in particular, for $N$-body simulations
  that are modified to be consistent with the general-relativistic constraints
  imposed by scalar averaging
  \postrefereeAAchanges{(Roukema et al., in preparation)}
  and for other simulational approaches working towards
    general-relativistic cosmology
    (\citealt{GiblinMS15departFLRW,GiblinMS16raytrace,BentivegnaBruni15,AdamekDDK15};
    \postrefereeAAchanges{\citealt{RaczDobos16,Daverio17,Macpherson17}}).

One of the main proposals for a relativistic improvement over
$\Lambda$CDM is the scalar averaging approach
\citep{Buch00scalav,Buch01scalav,Buchert11Towards}, which, in general,
is {\em background-free}. This approach extends the
Friedmann and acceleration equations (Hamiltonian constraint and Raychaudhuri equation)
from the homogeneous case
to general-relativistically take into account inhomogeneous curvature
and inhomogeneous expansion of the Universe
\citep{Rasanen04,Buchert08status,WiegBuch10,BuchRas12}.
\prerefereeCCchanges{This leads to a candidate explanation of
dark energy being the recent emergence of average negative scalar
curvature
\citep{Buchert05CQGLetter},
in particular by
dividing the spatial section into complementary
under- and overdense regions
(\citealt{BuchCarf08, Buchert08status};
for a related phenomenological lapse function
approach, see \citealt{Wiltshire07clocks, Wiltshire07exact})}.
Deviations of the average curvature from a constant-curvature model
are induced by kinematical backreaction, {together}
obeying a combined conservation law \citep{Buch00scalav}, while
implying global gravitational instability of the FLRW model and
driving the average model into the dark energy sector on large scales
\citep{RoyFOCUS}.

In practice, even when developing a background-free implementation of a scalar-averaged
cosmological model, an EdS model still provides the
simplest choice for initial conditions, so that the question of choosing
an observationally acceptable value of $\Hzerobgifbgtrue$ arises.
Existing implementations of emerging average negative
curvature models, include, {e.g.},
\prerefereeCCchanges{toy models of
  collapsing and expanding spheres
  \citep{Rasanen06negemergcurv} or
  Lema\^{\i}tre--Tolman--Bondi (LTB) regions
  \citep{NambuTanimoto05,KaiNambu07},
  a} peak model
\citep{Rasanen08peakmodel}, a metric template model
{\citep{Larena09template,Chiesa14Larenatest},} bi-scale or {more
  general} multi-scale models \citep{WiegBuch10,BuchRas12}, the
Timescape model \citep*{Wiltshire09timescape,DuleyWilt13,NazerW15CMB},
the virialisation approximation \citep*{ROB13},
{an effective viscous pressure approach \citep{Barbosa16viscos},}
and
Swiss cheese models that paste exact
inhomogeneous solutions into holes in a homogeneous (FLRW) background
(\citealt*{BolejkoC10SwissSzek}; the Tardis model of \citealt*{LRasSzybka13}).
Updates to many of these models should benefit from an observationally justified
estimate of $\Hzerobgifbgtrue$.
[See also recent work
  on averaging of
  LTB \citep{Sussman15LTBpertGIC,Chirinos16LTBav}
  and Szekeres models \citep{Bolejko08Szekaverage};
  for evolving sign-of-curvature models,
  see e.g. \citet{Kras81kevolving,Kras82kevolving,Kras83kevolving,Stichel16};
  \postrefereeAAchanges{for averaging using Cartan scalars,
    see \citet{Coley10scalars,KasparSvitek14Cartan}}.]

Moreover,
the ratio $\Hzerobg/\Hzeroeff$, where
$\Hzerobg$ is the present value of the background EdS model Hubble parameter and
  $\Hzeroeff$ is the effective low-redshift Hubble constant
(\citealt{Lemaitre31ell}; see also \citealt{Hubble1929}),
is another key property {that should emerge in a relativistic cosmological model.}
If this ratio is as small as $\Hzerobg/\Hzeroeff \sim 1/2$ \citep[cf][]{ROB13},
then, through Eqs.~(\ref{e-omm-H0-Hbg0}) and
(\ref{e-curv-eqn}), presented below in \SSS\ref{s-key-eqns},
an {\em observational} order of unity effect on the
effective density and curvature parameters is expected in comparison
to their values in a decoupled (FLRW) model, so that average recent-epoch
hyperbolicity (negative curvature) can provide the main component of
``dark energy''.  This responds to the commonly raised objection to
dark-energy--free scalar averaged models, according to which the
theoretically expected emergent average negative curvature is of an
order of magnitude too small to explain the needed amount of dark
energy, e.g.  in the
{conservative}\footnote{{\citet{BuchRZA2}
    use a generic scalar-averaging formalism, but implement it
    in a
    {background-dependent} way.}}
{approach
  of \citet*{BuchRZA2},} 
where
the
{overall} backreaction magnitude is found to
{lie in the range of a few percent}
on large
scales.

Contrary to the popular conception that spatial curvature is
tightly constrained observationally, observational constraints on
recently emerged, {\em present-day} average negative curvature
\postrefereeAAchanges{(denoted $\OmRzeroeff$ in Eq.~(\ref{e-curv-eqn})
  below)}
are
weak.  For example, \citet{Larena09template} and \citet{SaponeMN14}
applied the Clarkson, {Bassett} \& Lu test
\citep{ClarksonBL08test,Clarkson12test}, but found that existing
catalogues are not yet accurate enough.  Curvature constraints that
use cosmic microwave background (CMB)
\postrefereeAAchanges{and/or} baryon acoustic oscillation
(BAO) data and assume an FLRW model are {\em precise} in estimating
the homogeneous curvature parameter $\Omkzero$ to be bound by
$|\Omkzero| \ltapprox 0.005$ {(\citealt{Planck2015cosmparam}; or,
  e.g., $|\Omkzero| \ltapprox 0.009$, \citealt{ChenRatraBLZ15}).}
However, they are {\em inaccurate} in the sense that they do not allow
for average comoving curvature evolution when fitting the
observational data,
\postrefereeAAchanges{i.e. $\Omkzero$ is unlikely to be a good approximation
  to $\OmRzeroeff$.}
\postrefereeAAchanges{This restriction leads to inaccuracy} because voids, which dominate
the volume of the recent Universe, are general-relativistically
characterised by an average negative scalar curvature. The latter
effect is mirrored by, for example, the recent growth of the
virialisation fraction, which is a dimensionless parameter that can be
used to measure inhomogeneity growth for the complementary over-dense
structures \citep{ROB13}.

The details of individual effective models vary.
Here, we use the bi-scale scalar averaging approach
{\citep[e.g.,][and references therein]{ROB13}}.
In \SSS\ref{s-method} we summarise our assumptions
(\SSS\ref{s-assumptions}) and
{present the key equations}
(\SSS\ref{s-key-eqns}).
These provide relations among
{five} present-epoch cosmological
parameters and one {early-epoch--normalised
cosmological parameter,} $\Hzerobgifbgtrue$.
In \SSS\ref{s-lcdm-proxy} we use some properties of the $\Lambda$CDM model,
considered as an observational proxy, to derive {estimates of
  $\Hzerobgifbgtrue$ and $\Hzerobg$.}
Since low-redshift observational properties of the $\Lambda$CDM proxy are
primarily spatial, not temporal,
we discuss the consequences for the age of the {Universe}
in \SSS\ref{s-t0}.
We quantify the challenge in {estimating} recently-emerged curvature
in \SSS\ref{s-avcurv-howto}.
{We conclude in \SSS\ref{s-conclu}.}




\section{Implementation of scalar averaging} \label{s-method}

{We aim here to make a minimal number
of assumptions. While implementations of scalar averaging and other relativistic approaches
to cosmology vary, these assumptions are generally adopted, even if implicitly.}

\subsection{Model assumptions} \label{s-assumptions}

{As in several scalar averaging {implementations,}
  such as that} of \cite{ROB13},
we assume:
\begin{enumerate}[label=\protect\postrefereeAAchanges{(\roman*)}]
  \item
  an Einstein--de~Sitter (EdS) ``background'' model at early times,
  which we extrapolate
  to the present;
  the model is parametrised by
  {$\Hzerobgifbgtrue := \Hbg(\abg =1)$,
    where the background scale factor
    {$\abg$
      and Hubble parameter $\Hbg$} are given by
    \begin{align}
      \abg := (3 \Hzerobgifbgtrue t /2)^{2/3} \,, \quad
      \Hbg := \dotabg/\abg = 2/(3t)
      \,,
      \label{e-defn-Hbg1}
  \end{align}}
  {and the effective scale factor $\aeff$ (normalised to
    $\aeff = 1$ at the present time $\tzeroeff \equiv \tzeroefffull$) satisfies
    $\aeff \approx \abg$ at early times;}
  \label{assump:EdS}
\item
  zero cosmological constant/dark energy, i.e. $\Lambda:=0$;\label{assump:DE-free}
\item
  {bi-domain scalar}
  averaging---division of a spatial slice into over- and underdense
  regions; and\label{assump:bidomain}
\item
  virialisation of
  collapsed (overdense)
  regions,
  {i.e. these are assumed to have
    a negligible expansion rate}
  {(stable clustering in real space,
    e.g.,
    \citealt{Peebles1980,Jing01stableclust})};\label{assump:stableclustering}
  \end{enumerate}
{and we} define
\begin{align}
  \Hzerobg := \Hbg(\aeff=1)\,.
  \label{e-defn-Hbg0}
\end{align}
We refer to
scalar averages, denoted ``eff'', as ``parameters'', i.e. for a
{fixed large scale of statistical homogeneity
  \citep[e.g.~][]{HoggEis05,Scrimgeour12WDEShomog,WiegBO14}}.
The EdS high-redshift assumption
\ref{assump:EdS}
is observationally
realistic.
{Although $\Ommzero$
is often written as $\Omm$} for convenience,
$\Omm(z)$ in the FLRW models is (in general) $z$-dependent.
{In $\Lambda$CDM, $\Omm(z = 1100) \approx 1 - 10^{-9}$
(ignoring energy density components such as radiation and neutrinos),}
which is
observationally indistinguishable at that redshift
from the EdS value of $\Omm(z \approx 1100) = 1$.

\subsection{Key equations} \label{s-key-eqns}

{Since the spherical collapse overdensity threshold is several hundred,
  {volume-weighted averaging,
    together with assumptions} \ref{assump:bidomain} and \ref{assump:stableclustering}, imply that
  the average expansion rate is
  {close to}
  that of the underdense regions,
  {especially at late times,}
  and can be rewritten as}
\begin{align}
  \Heff(t) \approx \Hbg(t) + \Hpeculiar(t)
  \label{e-expansion-eqn-general},
\end{align}
where $\Hpeculiar$ is the peculiar expansion rate of voids,
i.e. the expansion rate above that of the extrapolated high-redshift
background EdS model
{(cf.\/ eq.~(32) of \citealt{BuchCarf08}; eq.~(2.27) of \citealt{ROB13})}.
At early epochs, prior to the main
virialisation epoch,
the expansion is dominated by the EdS background model, i.e.,
\begin{align}
  \Heff \approx \Hbg { = \Hzerobgifbgtrue \abg^{-3/2}}\,,
  \label{e-Heff-high-z}
\end{align}
while at the present, the effective local expansion
\postrefereeAAchanges{(measured by local estimates of the Hubble constant)},
is the sum of the
background expansion rate and the peculiar expansion rate of voids, i.e.,
\begin{align}
  \Hzeroeff \approx \Hzerobg + \Hpeculiarzero \,.
  \label{e-expansion-eqn-now}
\end{align}

Since we have an early epoch EdS model that we extrapolate to later epochs (assumption \ref{assump:EdS}),
{matter conservation gives the effective present-day matter density parameter
  \citep[e.g., eq.~(6),][]{BuchCarf08}
  \begin{align}
    \Ommzeroeff
    &= \frac{\Ommzerobg}{\left(\Hzeroeff/\Hzerobg\right)^2}
    \left(\frac{\abgzero}{\aeffzero}\right)^3
    = \abgzero^3 \left(\frac{\Hzerobg}{\Hzeroeff}\right)^2
    {,}
    \label{e-ommzeroeff}
  \end{align}
  where $\abgzero, \aeffzero$ are the present values of $\abg, \aeff$, respectively.
Equation~(\ref{e-ommzeroeff})
has solutions
$ \Hzerobg {=} \pm \Hzeroeff \sqrt{\Ommzeroeff /\abgzero^3}.$} A
high-redshift ($z \gtapprox 3$)
model that contracts
would not be physically realistic, so we have
positive $\Hzerobg$. We have a void-dominated model,
so we also have positive $\Hpeculiarzero$. Thus, the solution of
{physical interest is}
\begin{equation}
  \Hzerobg {=}
  \Hzeroeff \sqrt{\Ommzeroeff {/\abgzero^3}}
  {\,.}
  \label{e-omm-H0-Hbg0}
\end{equation}
We can now estimate the effective scalar curvature.
{The
  Hamiltonian constraint
{[e.g., eq.~(7), \citealt{BuchCarf08}] }
at the present epoch gives}
\begin{equation}
  \OmRzeroeff {=} 1 - \Ommzeroeff {- \OmQzeroeff} \;,
  \label{e-curv-eqn}
\end{equation}
where $\OmRzeroeff$ is the effective (averaged) present-day
scalar (3-Ricci) curvature
parameter
and $\OmQzeroeff$ is the
effective (averaged)
  present-day kinematical
  backreaction parameter\footnote{\protect\postrefereeAAchangesnosize{See, e.g., eq.~(2.9), \cite{ROB13}.
      The sum $\Omega_{\CX} := \OmReff + \OmQeff$
      \citep[Sect.~2.4.1,][]{Buchert08status}
      is not only a relativistic alternative to
      dark energy on large scales, it may also provide a relativistic contribution
      to dark matter on small scales.}}.
As summarised in \SSS\ref{s-intro},
observational constraints on recently-evolved average spatial hyberbolicity
remain weak, and we comment on this further in \SSS\ref{s-avcurv-howto}.

The FLRW equivalents of
two of the parameters in Eq.~(\ref{e-omm-H0-Hbg0})---$\Hzeroeff$ and
$\Ommzeroeff$---have been the subject of low-redshift observational
work for many decades.
{In Eqs~(\ref{e-Hzerobg-howto}) and (\ref{e-omm-H0-Hbg1})
  below, we show that adding a third long-studied observational parameter,
  $\tzeroeff$,
  lets us observationally estimate both
  $\Hzerobgifbgtrue$ and $\Hzerobg$
  if we use the $\Lambda$CDM
  model as a {\em proxy}, in the sense that
  it provides a phenomenological fit to many observations.}


\section{$\Lambda$CDM as an observational proxy} \label{s-lcdm-proxy}

{With the aim of using $\Lambda$CDM as an extragalactic
  observational proxy, we can use Eq.~(\ref{e-defn-Hbg1}) and assumption \ref{assump:EdS} to
  write $\Hzerobg$ in alternative form to that in
  Eq.~(\ref{e-omm-H0-Hbg0}), i.e.,
  \begin{align}
    \Hzerobg = 2/(3\tzeroeff)\,,
    \label{e-Hzerobg-howto}
  \end{align}
  and using
  Eqs~(\ref{e-defn-Hbg1}) and
  (\ref{e-omm-H0-Hbg0}) we can now write $\Hzerobgifbgtrue$ as
  \begin{align}
    \Hzerobgifbgtrue = \Hzeroeff \sqrt{\Ommzeroeff}\,.
    \label{e-omm-H0-Hbg1}
  \end{align}
  Thus, Eqs~(\ref{e-Hzerobg-howto}) and (\ref{e-omm-H0-Hbg1}) show that
  $\Hzerobg$ and
  $\Hzerobgifbgtrue$ are constrained by the values of
  $\Ommzeroeff$, $\Hzeroeff$ and $\tzeroeff$,
  estimated either by methods that minimise model dependence, or
  by using $\Lambda$CDM as an observational proxy for these values.}

Ideally, moderate-$z$ non-CMB
{$\Lambda$CDM proxy
estimates of $\Ommzeroeff$ and $\Hzeroeff$
should be
used} in Eq.~(\ref{e-omm-H0-Hbg1}) in order to estimate $\Hzerobgifbgtrue$.
For example, fitting the FLRW $H(z)$ relation at moderate redshifts
(e.g. $0.1 \ltapprox z \ltapprox 10$)
determined by differential oldest-passive-galaxy
stellar-population age
dating (``cosmic chronometers''; \citealt{JimLoeb02cosmchronom})
and using the FLRW fitted values of $\Ommzeroeff$ and $\Hzeroeff$
would be an observational strategy with only weak FLRW model dependence,
{especially if the technique became viable with $z \gtapprox 3$ galaxies.
For galaxies with $z < 3$,
some} authors find no significant inconsistency with $\Lambda$CDM
\citep[e.g.][]{Moresco16cosmchron}, while others find
tentative evidence for a non-$\Lambda$CDM $H(z)$ relation
(\citealt{DingBiesiada15Om}; see also the BAO estimates of
\citealt{SahniShSt14Om}).
Here, our main aim is to illustrate our method, so for simplicity,
we adopt $\Lambda$CDM as a proxy for a wide (though not complete,
e.g., \citealt{Bull15BeyondLCDM,BCKRW16}) range of extragalactic observations.
This should provide a reasonable initial estimate of $\Hzerobgifbgtrue$.
Adopting Planck values of $\Ommzero = 0.309 \pm 0.006, H_0 = 67.74 \pm 0.46${\kmsMpc}
\citep[Table 4, 6th data column,][]{Planck2015cosmparam},
Eq.~(\ref{e-omm-H0-Hbg1})
gives
\begin{align}
  \Hzerobgifbgtrue = {37.7 \pm 0.4}{\kmsMpc},
  \label{e-Hzerobgifbgtrue-LCDM-2015}
\end{align}
where, for the sake of illustration, the errors
in the $\Ommzero$ and $H_0$ estimates are assumed to be gaussian
and independent, with zero covariance.\footnote{The recent
  discovery
  {of differential space expansion
    on the hundred-megaparsec scale around our Galaxy
    (\citealt*{BolNazWilt16}; see also \citealt{KraljicSarkar16unifHubble})
    and the percent-level deviation of the
    \citet{Riess16H73}
    estimate of $\Hzeroeff$
    from the Planck estimate indicate percent-level effects on $\Hzeroeff$ when
    averaging on the hundred-megaparsec scale
    (see also \citealt{BDDurrerMSchwarz14}), which would modify the
    estimates presented here
    {at a similarly weak level.}}}
\prerefereeAAchanges{After submission of the present paper, we found that
  \cite{RaczDobos16} derived an almost identical $\Lambda$CDM-proxy value.}

Similarly,
{Eq.~(\ref{e-Hzerobg-howto}),
  using the Planck age of the Universe estimate $\tzeroLCDM = 13.80\pm0.02$~Gyr
  as a proxy,}
yields
\begin{align}
  \Hzerobg = 47.24 \pm 0.07{\kmsMpc}.
  \label{e-Hzerobg-estimate}
\end{align}
\prerefereeBBchanges{This is significantly higher than
  direct EdS fits to the CMB with broken-power-law or bump primordial spectra
  \citep{BlanchSarkar03,HuntSarkar07glitches,NadSarkar11LTB}, e.g.
  $43.3\pm 0.9${\kmsMpc} for what in our terminology appears to
  correspond to $\Hzerobg$
  \citep[][Table~2]{HuntSarkar10}.}

{Comparison of
  Eqs.~(\ref{e-omm-H0-Hbg0}) and (\ref{e-omm-H0-Hbg1}) gives
  the corresponding present-day background scale factor
  \begin{align}
    \abgzero =
    \prerefereeAAchanges{\left({\Hzerobgifbgtrue}/{\Hzerobg}\right)^{2/3} =
      0.860 \pm 0.007}
  \,, \label{e-abgzero}\end{align}
  which is
  \prerefereeCCchanges{slightly stronger than
  the $\approx\,$10\%} shrinkage in the BAO peak location detected for
  Sloan Digital Sky Survey 
  Luminous Red Galaxy 
  pairs whose paths cross superclusters
  \postrefereeAAchanges{in either the
  \citet{NadHot2013} or \citet{Liivamagi12} supercluster catalogues
  \citep*{RBOF15,RBFO15}.
  This suggests that BAO-peak--scale regions crossing superclusters could be considered as
  a slightly expanded present-day physical realisation of the
  EdS model extrapolated from early epochs, which we refer to
  in this paper as our background model.
  The values discussed below in \SSS\ref{s-t0} yield
  $\abgzero = 0.90 \pm 0.01$, in which case the EdS background model
  and the BAO-peak--scale regions
  crossing superclusters correspond even more closely.}

\section{Astrophysical age of universe estimates as a test of
  {inhomogeneous} cosmology}
\label{s-t0}

The value of $\Hzerobgifbgtrue$ in Eq.~(\ref{e-Hzerobgifbgtrue-LCDM-2015})
gives
$\tzerobg = 17.3$~Gyr [cf. Eq.~(\ref{e-defn-Hbg1})].  In a scalar
averaging model, {$\aeff(t) > \abg(t)$ (and $\Heff(t) > \Hbg(t)$)} are
expected, especially during the structure formation epoch, so the
expected present age of the Universe should be lower, i.e.
{$\tzeroeff < 17.3\,\mathrm{Gyr}.$} A model that provides $\aeff = 1$
at 13.8~Gyr would closely match $\Lambda$CDM.  However, by evolving an
initial power spectrum of density perturbations from an early epoch
forward in foliation time,
\postrefereeAAchanges{predictions} of
$\tzeroeff$ that differ from the $\Lambda$CDM value can also be made.
\postrefereeAAchanges{For example, this evolution can be calculated
  using the
  relativistic Zel'dovich approximation \citep{Kasai95,Kasai98} in the form
  given by \citet{BuchRZA1,BuchRZA2,BuchRZA3}; see also
  \citet{MatarreseTerranova96,VillaMatarrese11}.}

Use of $\Lambda$CDM as a proxy in \SSS\ref{s-lcdm-proxy} can
be considered to be approximately calibrated by differential passive galaxy age dating,
which relates
the effective scale factor and the time foliation,
at redshifts $0.1 \ltapprox z \ltapprox 2$.
However, at low
redshifts, observational constraints on $\Lambda$CDM mostly do not directly
relate to the time foliation. For example, ``observed peculiar velocities''
are combinations of spectroscopic redshifts, distance estimators, and
an assumed value of $H_0$; they are not measured spatial displacements differentiated
with respect to measured foliation time.
{Moreover, Ly$\alpha$ BAO estimates
  for $\Heff(z \sim 2.34)$ in the radial direction
  are about 7\% lower than the $\Lambda$CDM
  expected value \citep{Delubac14BAOLyaforest}, suggesting
  an underestimate of similar magnitude when using $\Lambda$CDM as a proxy
  to estimate $\tzeroeff$.}
In other words, it is premature to claim that {$\tzeroeff$}
{is} {accurately estimated to within $\pm$0.1~Gyr by} $\tzeroLCDM = 13.8$~Gyr.

  \begin{figure}
    \input{t_max_age}
    \caption{{Skew-normal reconstructed
        [Eq.~(\ref{e-skew-normal})]
        probability density functions $p_i$ of the twelve
      \protect\citet{Bensby13Bulgetzero} microlensed Galactic Bulge stars
      whose most likely age is greater than 13.0~Gyr
      (thin curves); and probability density function
      $\diffd P_T/\diffd t'$ of the most likely oldest age $T$ of these
      stars (thick curve),
      assuming that the stars' true ages are chosen randomly from their
      respective pdfs $p_i$
      [Eq.~(\protect\ref{e-P-T-defn})]. The vertical line indicates
      $\tzeroLCDM$.}
      \label{f-t-max-age}}
\end{figure}

How well is $\tzeroeff$ observationally constrained?
Here, {we} consider the integral of proper time on
{our Galaxy's world line}
from the initial singularity to the present
{to be negligibly different from the corresponding
  time {interval}
  in terms of coordinate time $t$,
  so that both can be consistently
  denoted by $\tzeroeff$
(see, however, \citealt{NazerW15CMB}, and references therein).}
Six microlensed bulge $\sim$1$\,M_{\odot}$ stars
have most-probable age estimates in the range 14.2--14.7~Gyr
\citep[table 5,][]{Bensby13Bulgetzero}. The
probability density functions (pdfs) of the age estimates
for these stars are highly asymmetric, with
{68\% confidence intervals}
each including $\tzeroLCDM$ and having
an upper bound below or nearly below the mode of the pdf.

A more precise estimate of a lower bound to
{$\tzeroGal$} can be obtained
if we interpret the pdfs of these microlensed stars to represent the present
best state of our knowledge of the ages of these stars. In that case,
we can remodel the published distributions using skew-normal distributions,
{i.e., with pdfs
\begin{align}
  p(t;\alpha,\xi,\omega) &= 
  {\frac{e^{-\tau^2/2}}{\omega \sqrt{2\pi}}
    \left[1 + \mathop{\mathrm{erf}}\left( \frac{\alpha \tau}{\sqrt{2}} \right) \right]} \,,
  \label{e-skew-normal}
\end{align}
where}
${\tau :=} (t-\xi)/\omega$ is a rescaled age,
$\xi$ is a location parameter,
$\omega$ is a width parameter,
$\alpha$ is an asymmetry parameter
\citep{deHelguero1909,Birnbaum50,Azzalini05skewnormal}. We numerically search
for (using simulated annealing)
the triple $(\alpha,\xi,\omega)$ that best matches the age
pdf parameters in columns 5, 6, and 7
of Table~5 of \citet{Bensby13Bulgetzero} for each of the stars whose most likely
age (column 5) is greater than 13.0~Gyr, yielding $p_i$ for star $i = 1, \ldots, 12$
satisfying this definition of old stars.
The cumulative probability {$P_T$ that none of the 12 stars is
older than an age $t'$ is}
\begin{align}
  {P_{T}
  (T \le t' )
    = \Pi_{i=1,12} \int_{-\infty}^{t'}  p_i(t) \,\diffd t.}
  \label{e-P-T-defn}
\end{align}
{As shown in
  Fig.~\ref{f-t-max-age},
this} gives the most likely {value for $T$, the age of the oldest star in this sample, as}
$T = {14.7^{+0.3}_{-0.7}}\,\mathrm{Gyr}$
{({the uncertainties indicate the
  68\% central} confidence interval in $P_T$; or
   $ 14.7^{+0.8}_{-1.5}\,\mathrm{Gyr}$, at 95\%).}
This value provides a lower bound for $\tzeroGal$; assuming this oldest star formed very early in the Universe, i.e. $T \approx \tzeroGal$, would give
$\Hzerobg = 44^{+1}_{-2}${\kmsMpc},
i.e.,
\postrefereeAAchanges{about 3{\kmsMpc} lower}
than the value in Eq.~(\ref{e-Hzerobg-estimate}),
\prerefereeBBchanges{and consistent with
  \citet{HuntSarkar10}'s CMB estimate of
  $43.3\pm 0.9${\kmsMpc} quoted above. The latter type of estimate appears to
  be convergent with respect to increasing data quality and improved analysis
  \citep{BlanchSarkar03,HuntSarkar07glitches,NadSarkar11LTB}.}
\postrefereeAAchanges{As mentioned above, this lower estimate of
  $\Hzerobg$ yields $\abgzero = 0.90 \pm 0.01$,
  improving the correspondence between the BAO peak shift
  (shrinkage across superclusters; \citealt{RBOF15,RBFO15})
  and $\abgzero$.}

With improved stellar modelling accuracy, estimates of this sort can potentially be used to distinguish $\Lambda$CDM
from relativistic
{inhomogeneous}
models. However, the present derivation of {$P_{T}$} by
analysing the \citet{Bensby13Bulgetzero} microlensed
stars is {\em not} sufficient to reject $\Lambda$CDM. Firstly, the probability
of {having $T \leq \tzeroLCDM$ according to $P_{T}$} is 9\%, which is not a statistically significant
rejection. Secondly, and
more importantly, {$P_{T}$ as defined here is designed to provide the pdf of a best estimate
of a lower bound to $t_0$,} given the published modelling of the observational data and assuming that
the individual stellar pdfs are statistically independent; it is not
designed to test the hypothesis that $\tzeroLCDM$ is the true value of
{$\tzeroGal$}.
Thirdly, the typical uncertainties in stellar age modelling represented
in the pdfs $p_i$ are high. For example, fig.~12 of \citet{Dotter07} shows
variation in age of up to about $\pm 15\%$ if oxygen and iron abundances are
0.3~dex higher than what the authors refer to as ``scaled-solar'' abundances;
and figs~15 and 16 of \citet{VandenBerg12abundances} indicate that a 0.4~dex
enhancement in magnesium or silicon abundance has a stronger effect on
effective temperature $T_{\mathrm{eff}}$ than 2~Gyr in age.
In particular, \citet{VandenBerg14HD140283} estimate the age of the halo subgiant HD~140283
as $14.3\pm0.4$~Gyr, where the error is predominantly parallax error, or
$14.3\pm0.8$~Gyr, including all sources of uncertainty, such as
{that of} the oxygen abundance.

\section{Present-day/recent average curvature}
\label{s-avcurv-howto}

As stated above,
the present-day
{large-scale mean}
curvature represented in Eq.~(\ref{e-curv-eqn})
is not easy to measure
\citep[e.g.,][]{Larena09template,SaponeMN14,RasanenBF15distsum},
{but} will become measurable in near-future surveys such as
\postrefereeAAchanges{Euclid
  \citep{EuclidScienceBook2010}, DESI (Dark Energy Spectroscopic
Instrument; \citealt{Levi13DESI}), 4MOST (4-metre Multi-Object
Spectroscopic Telescope;
\citealt{deJong12VISTA4MOST,deJong14subscriponly4MOST}),
eBOSS
(extended Baryon Oscillation Spectroscopic Survey;
\citealt{Zhao15eBOSSpredict}), LSST (Large Synoptic Survey
Telescope; \citealt{TysonLSST03}), and
HETDEX (Hobby--Eberly Telescope Dark Energy eXperiment; \citealt{Hill08HETDEX}).}
{Power-law models of
  {dark-energy--free [cf. \ref{assump:DE-free}]}
  backreaction evolution tend
  to give {a strong} negative mean curvature (positive $\OmRzeroeff$),
  e.g. \citet[{\SSS}IV,][]{WiegBuch10} argue for $\OmRzeroeff \approx 1.05$,
  $\OmQzeroeff \approx -0.35$.
  For the purposes of illustration,
  we set $\OmQzeroeff = 0$ in this paragraph,
  giving a slightly weaker negative mean present-day curvature, with
  $\OmRzeroeff \approx 0.7$ from Eq.~(\ref{e-curv-eqn}).}
If a volume with $\OmRzeroeff \approx 0.7$ is
represented as a single time-slice
constant-curvature (``template'') model
(\citealt{BuchCarf02}; or alternatively, by smoothing the curvature
and neglecting the ``dressing'' of Riemannian volume and the subdominant
curvature fluctuation backreaction that both arise through the smoothing process;
see \citealt{BuchCarf03nudity}), then we can write an effective curvature radius
$\RCzeroeff \approx (c/\Hzeroeff) {\OmRzeroeff}^{-1/2} \approx
3580$ {\hMpc}.  Still assuming constant curvature, the difference
between the expected tangential arclength subtending a radian
$r_\perp^-$ and a flat space arclength $r_\perp^0$ at a small
radial comoving distance $r$ is $r_\perp^- - r_\perp^0 =
\RCzeroeff \sinh(r/\RCzeroeff) - r  \approx (1/6) r (r/\RCzeroeff)^2$
to highest order. So up to 300{\hMpc} from the observer, the
tangential ``stretch'' is less than about 0.1\%, i.e. BAO
curvature constraints would have to be accurate to better than
$\pm 0.1${\hMpc}. At a redshift $z=1$, the tangential effect
should be stronger, but in the past light cone, the {\em
  present-day} averaged curvature does not apply.  Instead, using
$\Lambda$CDM as a proxy, we should have
$\OmReff(z=1) \approx \Omega_\Lambda(z=1) = 0.23$, i.e.  average
negative curvature is weaker, with a
constant-curvature
curvature radius
\footnote{\protect\postrefereeAAchangesnosize{We use the adjective
    ``constant-curvature'' in front of ``curvature radius''
    to emphasise that interpreting the mean
    spatial curvature in terms of a curvature radius is only
    meaningful for some types of approximate calculations,
    such as for small angles.}}
$\RCeff(z=1) \approx 6310${\hMpc},
double the local value.  So in a constant cosmological time,
constant-curvature hypersurface at $z=1$,
\postrefereeAAchanges{the amount of}
tangential stretching \postrefereeAAchanges{that}
should occur at 500{\hMpc} from the observer
\postrefereeAAchanges{is about 0.1\%}.  At the BAO scale of
about $105${\hMpc}, stretching would {be} about 25 times weaker. It
remains to be determined whether $\sim 0.004\%$ stretching will be
detectable in the coming decade of major observational projects such
as \postrefereeAAchanges{Euclid, DESI, 4MOST, eBOSS, LSST, and HETDEX}.

\section{Conclusion} \label{s-conclu}
Equations
{(\ref{e-defn-Hbg1})--(\ref{e-omm-H0-Hbg1})
provide
a summary} of the key relationships between
present-day observational parameters of the
bi-domain scalar averaging model, satisfying
assumptions \ref{assump:EdS}--\ref{assump:stableclustering}.
The peculiar expansion rate of voids was previously estimated as
$\Hpeculiarzero \approx 36 \pm 3${\kmsMpc}
\citep[eq.~(2.36),][]{ROB13}
from corresponding surveys of galaxy clusters and voids,
{but
  Eqs~(\ref{e-expansion-eqn-now}) and (\ref{e-Hzerobg-estimate})
  imply that this is an overestimate by $\sim$10{\kmsMpc},
  even taking into account a somewhat greater $t_0$ motivated
  by stellar estimates.}
{Thus, following a scalar averaging (or similar) approach,}
$\Hpeculiarzero$,
together with the {early-epoch--normalised} background EdS Hubble constant $\Hzerobgifbgtrue$,
{the present background Hubble parameter $\Hzerobg$,}
the effective Hubble constant $\Hzeroeff$,
the effective
matter density parameter $\Ommzeroeff$,
{and the age of the Universe $\tzeroeff$}
form a closely linked {sextuple}.
Estimates of $\Hzeroeff$ and $\Ommzeroeff$ are generally uncontroversial,
though usually interpreted in terms of the gravitationally decoupled
(standard) cosmological {model.}

Here, we have shown that since $\Hzerobgifbgtrue$ is physically realised at high redshift rather
than low redshift, it can be given a value with a {reasonable}
observational justification, using $\Lambda$CDM as an observational proxy.
{The Planck $\Lambda$CDM values yield}
$\Hzerobgifbgtrue = {37.7 \pm 0.4}${\kmsMpc} [Eq.~(\ref{e-Hzerobgifbgtrue-LCDM-2015})].
{The} corresponding recently-emerged average negative spatial scalar curvature,
represented by the effective curvature
{parameter $\OmRzeroeff$,}
is not presently constrained by observations, and constitutes a key challenge
for observational cosmology in the coming decades.

{The corresponding high value of $\tzerobg = 17.3\,$Gyr
  motivates refocussing attention on astrophysical age estimates such as
  the microlensed oldest bulge star estimate
  ${T = 14.7^{+0.3}_{-0.7}}$~Gyr
  discussed above,
  since standard cosmological
  tools do not seem to provide CMB-free estimates of
  $\tzeroeff$.}
Modelling
of {suspected-oldest stars} with an appropriate statistical approach and
observational strategies could potentially
result in a {\em stellar} rejection of $\Lambda$CDM
\citep[cf.][]{VandenBerg14HD140283}.
As cosmological models with
standard (Einstein) gravity continue to be refined, predictions
of {$\tzeroeff$}
{will} need to be made and compared to improved stellar constraints.

Since the order of magnitude of cosmological backreaction effects is often
claimed to be tiny,
{the following}
order-of-magnitude summary of
{Eqs~(\ref{e-Heff-high-z})--(\ref{e-curv-eqn}) and (\ref{e-omm-H0-Hbg1})}
and their {values}
may help underline the
inaccuracy implied by ignoring standard Einsteinian
{gravity:
  \begin{align}
    \frac{2}{3}
    \approx \frac{\Hzerobg}{\Hzeroeff}
    \gtapprox \frac{\Hzerobgifbgtrue}{\Hzeroeff}
    \approx \sqrt{\Ommzeroeff}
    = \sqrt{1 - \OmRzeroeff - \OmQzeroeff}
    \approx \frac{1}{2}
    \gtapprox  \frac{\Hpeculiarzero}{\Hzeroeff}
    \,,
  \label{e-key-scalav-omegas}
  \end{align}
  {e.g.
    it is observationally realistic
    for the effective expansion rate to be as much as
    twice the
    background expansion rate.}
  Accepting} $\Hzeroeff$ and $\Ommzeroeff$
  as approximately well-known,
observational estimation of
{$\Hzerobg$ depends only on
  estimating $\tzeroeff$;
  $\Hzerobgifbgtrue$ is the main theme of
  this paper;
  $\OmRzeroeff$
  is only weakly constrained,
  although a strong positive value
  is expected due to the spatial dominance of voids; $\OmQzeroeff$
  is, in principle, measurable from distance--redshift catalogues;
  and $\Hpeculiarzero$}
  was estimated in \citet{ROB13} but remains open to improved methods.

\section*{Acknowledgments}
Thank you to
Roland Bacon,
Laurence Tresse,
Ariane Lan\c{c}on,
Johan Richard,
Subir Sarkar,
David Wiltshire,
Thomas Bensby
\postrefereeAAchanges{and an anonymous referee}
for useful comments.
  The work of TB and PM was conducted
    within the ``Lyon Institute of Origins'' under grant
    ANR-10-LABX-66.
  A part of this project was funded
  by the National Science Centre, Poland, under grant
  2014/13/B/ST9/00845.
  JJO benefited from doctoral stipendium financial support
  under grant DEC-2014/12/T/ST9/00100
  of the National Science Centre, Poland.
  Part of this work consists of
  research conducted within the scope of the HECOLS
  International Associated Laboratory, supported in part
  by the Polish NCN grant DEC-2013/08/M/ST9/00664. A part
  of this project has made use of computations made under
  grant 197 of the Pozna\'n Supercomputing and Networking
  Center (PSNC).

\subm{ \clearpage }

%



\end{document}

%% file: t_max_age.tex
\begingroup
  \makeatletter
  \providecommand\color[2][]{%
    \GenericError{(gnuplot) \space\space\space\@spaces}{%
      Package color not loaded in conjunction with
      terminal option `colourtext'%
    }{See the gnuplot documentation for explanation.%
    }{Either use 'blacktext' in gnuplot or load the package
      color.sty in LaTeX.}%
    \renewcommand\color[2][]{}%
  }%
  \providecommand\includegraphics[2][]{%
    \GenericError{(gnuplot) \space\space\space\@spaces}{%
      Package graphicx or graphics not loaded%
    }{See the gnuplot documentation for explanation.%
    }{The gnuplot epslatex terminal needs graphicx.sty or graphics.sty.}%
    \renewcommand\includegraphics[2][]{}%
  }%
  \providecommand\rotatebox[2]{#2}%
  \@ifundefined{ifGPcolor}{%
    \newif\ifGPcolor
    \GPcolorfalse
  }{}%
  \@ifundefined{ifGPblacktext}{%
    \newif\ifGPblacktext
    \GPblacktexttrue
  }{}%
  \let\gplgaddtomacro\g@addto@macro
  \gdef\gplbacktext{}%
  \gdef\gplfronttext{}%
  \makeatother
  \ifGPblacktext
    \def\colorrgb#1{}%
    \def\colorgray#1{}%
  \else
    \ifGPcolor
      \def\colorrgb#1{\color[rgb]{#1}}%
      \def\colorgray#1{\color[gray]{#1}}%
      \expandafter\def\csname LTw\endcsname{\color{white}}%
      \expandafter\def\csname LTb\endcsname{\color{black}}%
      \expandafter\def\csname LTa\endcsname{\color{black}}%
      \expandafter\def\csname LT0\endcsname{\color[rgb]{1,0,0}}%
      \expandafter\def\csname LT1\endcsname{\color[rgb]{0,1,0}}%
      \expandafter\def\csname LT2\endcsname{\color[rgb]{0,0,1}}%
      \expandafter\def\csname LT3\endcsname{\color[rgb]{1,0,1}}%
      \expandafter\def\csname LT4\endcsname{\color[rgb]{0,1,1}}%
      \expandafter\def\csname LT5\endcsname{\color[rgb]{1,1,0}}%
      \expandafter\def\csname LT6\endcsname{\color[rgb]{0,0,0}}%
      \expandafter\def\csname LT7\endcsname{\color[rgb]{1,0.3,0}}%
      \expandafter\def\csname LT8\endcsname{\color[rgb]{0.5,0.5,0.5}}%
    \else
      \def\colorrgb#1{\color{black}}%
      \def\colorgray#1{\color[gray]{#1}}%
      \expandafter\def\csname LTw\endcsname{\color{white}}%
      \expandafter\def\csname LTb\endcsname{\color{black}}%
      \expandafter\def\csname LTa\endcsname{\color{black}}%
      \expandafter\def\csname LT0\endcsname{\color{black}}%
      \expandafter\def\csname LT1\endcsname{\color{black}}%
      \expandafter\def\csname LT2\endcsname{\color{black}}%
      \expandafter\def\csname LT3\endcsname{\color{black}}%
      \expandafter\def\csname LT4\endcsname{\color{black}}%
      \expandafter\def\csname LT5\endcsname{\color{black}}%
      \expandafter\def\csname LT6\endcsname{\color{black}}%
      \expandafter\def\csname LT7\endcsname{\color{black}}%
      \expandafter\def\csname LT8\endcsname{\color{black}}%
    \fi
  \fi
  \setlength{\unitlength}{0.020bp}%
  \begin{picture}(11520.00,8640.00)%
    \gplgaddtomacro\gplbacktext{%
      \colorrgb{0.00,0.00,0.00}%
      \put(1480,1280){\makebox(0,0)[r]{\strut{}0}}%
      \colorrgb{0.00,0.00,0.00}%
      \put(1480,2656){\makebox(0,0)[r]{\strut{}0.2}}%
      \colorrgb{0.00,0.00,0.00}%
      \put(1480,4032){\makebox(0,0)[r]{\strut{}0.4}}%
      \colorrgb{0.00,0.00,0.00}%
      \put(1480,5407){\makebox(0,0)[r]{\strut{}0.6}}%
      \colorrgb{0.00,0.00,0.00}%
      \put(1480,6783){\makebox(0,0)[r]{\strut{}0.8}}%
      \colorrgb{0.00,0.00,0.00}%
      \put(1480,8159){\makebox(0,0)[r]{\strut{}1}}%
      \colorrgb{0.00,0.00,0.00}%
      \put(1720,880){\makebox(0,0){\strut{}13}}%
      \colorrgb{0.00,0.00,0.00}%
      \put(3233,880){\makebox(0,0){\strut{}13.5}}%
      \colorrgb{0.00,0.00,0.00}%
      \put(4746,880){\makebox(0,0){\strut{}14}}%
      \colorrgb{0.00,0.00,0.00}%
      \put(6260,880){\makebox(0,0){\strut{}14.5}}%
      \colorrgb{0.00,0.00,0.00}%
      \put(7773,880){\makebox(0,0){\strut{}15}}%
      \colorrgb{0.00,0.00,0.00}%
      \put(9286,880){\makebox(0,0){\strut{}15.5}}%
      \colorrgb{0.00,0.00,0.00}%
      \put(10799,880){\makebox(0,0){\strut{}16}}%
      \colorrgb{0.00,0.00,0.00}%
      \put(320,4719){\rotatebox{90}{\makebox(0,0){\strut{}probability
        density}}}%
      \colorrgb{0.00,0.00,0.00}%
      \put(6259,280){\makebox(0,0){\strut{}$t$ or $t'$ (Gyr)}}%
    }%
    \gplgaddtomacro\gplfronttext{%
      \colorrgb{0.00,0.00,0.00}%
      \put(10559,7846){\makebox(0,0)[r]{\strut{}${\diffd P_T/\diffd t'}$~}}%
      \colorrgb{0.00,0.00,0.00}%
      \put(10559,7346){\makebox(0,0)[r]{\strut{}$\tzeroLCDM$~}}%
      \colorrgb{0.00,0.00,0.00}%
      \put(10559,7096){\makebox(0,0)[r]{\strut{}$p_i$~~~~}}%
    }%
    \gplbacktext
    \put(0,0){\includegraphics[scale=0.4]{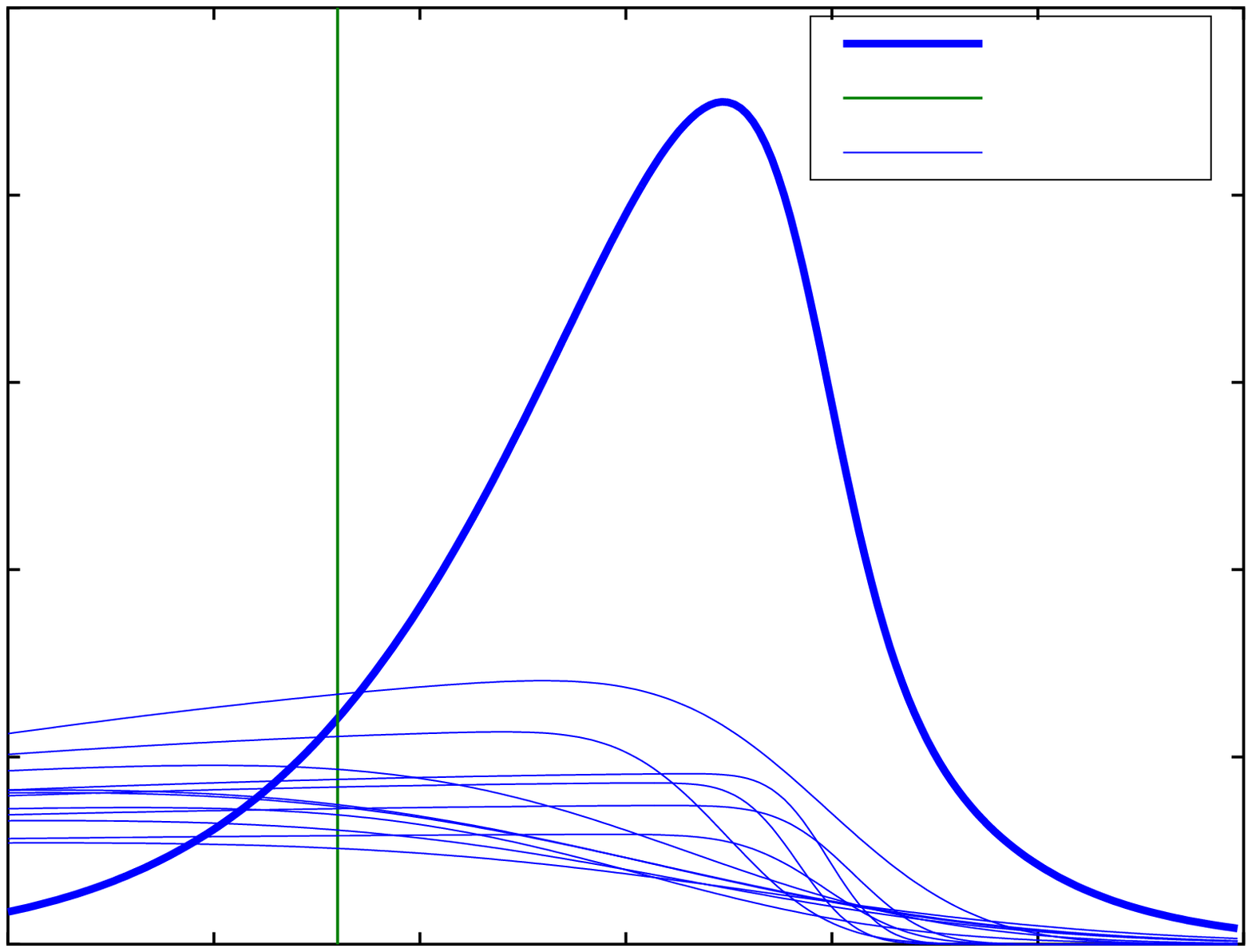}}%
    \gplfronttext
  \end{picture}%
\endgroup